\documentstyle[prl,aps,floats,epsfig]{revtex}  
\draft % makes pacs numbers print

\begin{document}
% YY-NN = your latest BN number 
%\preprint{\tighten\vbox{\hbox{\hfil Belle DRAFT {\it 01-476/481}G}
%                        \hbox{\hfil Intended for {\it Phys. Rev. Lett.}}
%               \hbox{\hfil Authors: F. Fang, M. Wang, 
% H.C Huang, T.E. Browder} 
%                        \hbox{\hfil Committee:A. Bondar, M. Nakao,
%                        S. Stanic}
%                        \hbox{\hfil TBA}
%                        \hbox{\hfil TBA}
%                        \hbox{\hfil TBA(chair)}
%}}

\title{\large
       Observation of $B^{\pm}\to p \bar{p} K^{\pm}$}

\maketitle

%\author{} 
% This is the B -> p pbar K paper's author list.
% Non-responding authors or those who said NO are commented out.
% ====================================================================
% Click the RELOAD button on your web browser to see the updated file.
% ====================================================================

\begin{center}
  K.~Abe$^{9}$,               % KEK
  K.~Abe$^{40}$,              % TohokuGakuin
% N.~Abe$^{43}$,              % TIT
  R.~Abe$^{30}$,              % Niigata
  T.~Abe$^{41}$,              % Tohoku
  I.~Adachi$^{9}$,            % KEK
  Byoung~Sup~Ahn$^{17}$,      % Korea
  H.~Aihara$^{42}$,           % Tokyo
  M.~Akatsu$^{23}$,           % Nagoya
% K.~Asai$^{24}$,             % Nara
% M.~Asai$^{10}$,             % Hiroshima
  Y.~Asano$^{47}$,            % Tsukuba
  T.~Aso$^{46}$
% ...
% IHEP
% Y.~Yusa$^{41}$,             % Tohoku
% H.~Yuta$^{1}$,              % Aomori
% C.~C.~Zhang$^{12}$,         % IHEP
  J.~Zhang$^{47}$,            % Tsukuba
% H.~W.~Zhao$^{9}$,           % KEK
  Y.~Zheng$^{8}$,             % Hawaii
% V.~Zhilich$^{2}$,           % BINP
and
  D.~\v Zontar$^{47}$         % Tsukuba
  \vskip1pc
  (Belle Collaboration)
\end{center}

\small
\begin{center}
% ...
$^{8}${University of Hawaii, Honolulu HI}\\
$^{9}${High Energy Accelerator Research Organization (KEK), Tsukuba}\\
% ...
$^{17}${Korea University, Seoul}\\
% ...
$^{23}${Nagoya University, Nagoya}\\
% ...
$^{30}${Niigata University, Niigata}\\
% ...
$^{40}${Tohoku Gakuin University, Tagajo}\\
$^{41}${Tohoku University, Sendai}\\
$^{42}${University of Tokyo, Tokyo}\\
% ...
$^{46}${Toyama National College of Maritime Technology, Toyama}\\
$^{47}${University of Tsukuba, Tsukuba}\\
% ...
%%%$^{\dagger}${on leave from University of Toronto, Toronto ON}
\end{center}

\normalsize
 
%\date{\today}

% to make it single spaced to save trees - preprint version
\tighten

\begin{abstract}

We report the observation of the decay mode 
$B^{\pm}\to p \bar{p} K^{\pm}$ based on an analysis of
$29.4$ fb$^{-1}$ of data collected by the Belle detector at
KEKB. This is the first example of a $b\to s$ transition
with baryons in the final state. 
The $p \bar{p}$ mass spectrum in this decay
is inconsistent with phase space and is peaked at low mass.
The branching fraction for this decay is measured to be
${\cal B}(B^{\pm}\to p \bar{p} K^{\pm})
 =(4.3^{+1.1}_{-0.9}({\rm stat})\pm 0.5({\rm syst}))\times 10^{-6}.$
We also report upper limits for the decays 
$B^0\to p \bar{p} K_S$ and $B^{\pm}\to p \bar{p} \pi^{\pm}$. 
%and $B^0\to p \bar{p} K^{*0}$ 
%but do not observe significant signals and report upper limits
%for these modes.
\vskip1pc

\end{abstract}
%\pacs{13.20.He, 13.25.Hw, 13.60.Rj}  

\twocolumn[\hsize\textwidth\columnwidth\hsize\csname
@twocolumnfalse\endcsname 
\normalsize
\vskip2pc ] 

\normalsize

We report the results of searches for the decay modes
$B^+\to  p \bar{p} K^+$\cite{CC} and 
$B^0\to p \bar{p} K_S$.
These modes are expected to proceed mainly via $b\to s$
penguin diagrams. We also search for
$B^{+}\to p \bar{p} \pi^{+}$ which is expected
to occur primarily via a $b\to u$ tree process. Once they are
established, these baryonic modes
may be used to either constrain or observe direct $CP$ violation 
in $B$ decay\cite{rosner_baryon}.

In contrast to charm meson decay, final states
with baryons are allowed in $B$ meson decay. 
To date, a few low multiplicity
$B$ decay modes with baryons in the final state from
$b\to c$ transitions have been observed\cite{cleobcbary}. 
Rare $B$ decays due to charmless
$b\to s$ and $b\to u$ transitions should also lead to final states
with baryons. A number of searches for such modes have been carried out
by CLEO\cite{cleorrbary}, ARGUS\cite{argusbary}, 
and LEP\cite{delphibary} but only upper limits were obtained. 
Stringent upper limits for two-body modes such as $B^0\to p \bar{p}$, 
$B^+\to \bar{\Lambda} p$ and  
$B^0\to \Lambda \bar{\Lambda}$ have recently been 
reported by Belle\cite{bellebary}.

We use a  $29.4~{\rm fb}^{-1}$ data sample,
which contains 31.9 million produced $B\bar{B}$ pairs, 
collected  with the Belle detector at the KEKB asymmetric-energy
$e^+e^-$ (3.5 on 8~GeV) collider~\cite{KEKB}.
KEKB operates at the $\Upsilon(4S)$ resonance 
($\sqrt{s}=10.58$~GeV) with a peak luminosity that exceeds
$5\times 10^{33}~{\rm cm}^{-2}{\rm s}^{-1}$.
The Belle detector is a large-solid-angle magnetic
spectrometer that
consists of a three-layer silicon vertex detector (SVD),
a 50-layer central drift chamber (CDC), a mosaic of
aerogel threshold \v{C}erenkov counters (ACC), time-of-flight
scintillation counters (TOF), and an array of CsI(Tl) crystals
(ECL)  located inside 
a superconducting solenoid coil that provides a 1.5~T
magnetic field.  An iron flux-return located outside of
the coil is instrumented to identify $K_L$ and
muons (KLM).  The detector
is described in detail elsewhere~\cite{Belle_nim}.

% add some particle id details

We select well measured
charged tracks with impact parameters with respect 
to the interaction point of less than 0.3 cm in the radial direction
and less than 3 cm in the beam direction ($z$). 
These tracks are required to
have $p_T>50$ MeV/$c$.

Particle identification likelihoods for each particle hypothesis
are calculated by combining
information from the TOF, ACC system with $dE/dx$ measurements in the CDC.
Protons and anti-protons are identified using all particle ID
systems and are required to have proton likelihood ratios 
($L_p/(L_p+L_K)$ and $L_p/(L_p+L_{\pi})$) greater than
0.6. Proton candidates that are electron-like according to the
information recorded by the CsI(Tl) calorimeter are vetoed.
This selection is $89\%$ efficient for protons
with a $7\%$ kaon misidentification rate.
To identify kaons (pions), we
require the kaon (pion) likelihood ratio
to be greater than 0.6. This requirement is 88\% efficient for
kaons with a $8.5\%$ misidentification rate for pions.
In addition, we remove kaon candidates that are consistent with being
protons. 

For the $B^0\to p \bar{p} K_S$ mode, we select $K_S$ candidates
from $\pi^+\pi^-$ candidates that lie within the mass window
$0.482 {\rm ~GeV}/c^2<M(\pi^+\pi^-)<0.514$ GeV/$c^2$ ($\pm 4 \sigma$). 
The distance of closest approach between the two
daughter tracks is required to be less than $2.4$ cm. 
The impact parameter of each track in the radial direction  should have
magnitude greater than $0.02$ cm, and the flight length
should be greater than 0.22 cm. The difference in the
angle between the 
vertex direction and the $K_S$ flight direction in the $x-y$ plane is
required to satisfy $\Delta\phi<0.03$ rad.

To reconstruct signal candidates in the $B^+\to  p \bar{p} K^+$ mode, 
we form combinations of a kaon, proton and anti-proton
that are inconsistent with the following $b\to c \bar{c} s$ transitions:
$B^+\to J/\psi K^+$, $J/\psi \to p \bar{p}$; $B^+\to \eta_c K^+$,
$\eta_c\to p \bar{p}$; $B^+\to \psi^{\prime} K^+$, 
$\psi^{\prime}\to p \bar{p}$
and $B^+\to \chi_{c[0,1]} K^+$, $\chi_{c[0,1]} \to p \bar{p}$.
This set of requirements is referred to as the charm veto\cite{vetos}.
Similar charm vetoes are applied in the analysis of
the other decay modes. In the case of $B^0\to p \bar{p} K_S$,
events with $p K_S$ or $\bar{p} K_S$ 
masses consistent with the $\Lambda_c$ are rejected.

To isolate the signal, we form the beam-constrained mass, 
 $M_{\rm bc}=\sqrt{E_{\rm beam}^2-\vec{P}_{\rm recon}^2}$, and
energy difference $\Delta E= E_{\rm recon}-E_{\rm beam}$
in the $\Upsilon(4S)$ center of mass frame. Here $E_{\rm beam}$,
$E_{\rm recon}$ and $\vec{P}_{\rm recon}$
are the beam energy, the reconstructed energy and the reconstructed
momentum of the signal candidate, respectively.
The signal
region for $\Delta E$ is $\pm 50$ MeV which corresponds to
$\pm 5\sigma$ where $\sigma$ is the resolution determined from a
Gaussian fit to the Monte Carlo (MC) simulation. 
The signal region for $M_{\rm bc}$ is $5.270 {\rm ~GeV}/c^2<M_{\rm bc}<5.290$
 GeV/$c^2$. The resolution
in beam-constrained mass is 2.8 MeV/$c^2$ and is dominated by the
beam energy spread of KEKB.

Several event topology variables provide discrimination
between the large 
continuum ($e^+e^-\to q \bar{q}$, where $q= u, d, s, c$) 
background, which  tends to be collimated along the original quark 
direction, and more spherical $B\bar{B}$ events.
%We first remove events with $R_2$, the second
%Fox-Wolfram moment, greater than 0.5. 
We form a likelihood
ratio using two variables. Six modified Fox-Wolfram
moments and the cosine of the thrust angle are combined into a Fisher 
discriminant\cite{SFW}. For signal MC and continuum data,
we then form probability density functions 
for this Fisher discriminant,
 and the cosine of
the $B$ decay angle with respect to the $z$ axis ($\cos\theta_B$).
The signal (background) 
probability density functions are multiplied together to form a signal
(background) likelihood ${\cal L}_S$ (${\cal L}_{BG})$. The
 likelihood ratio ${\cal L}_S/({\cal L}_S+{\cal L}_{BG})$ is then required
to be greater than 0.6. The event
topology requirements retain 78\% of the
signal while removing 87\% of the continuum background.

\begin{figure}[htb]
%\centerline{
%\epsfxsize 1.9 truein \epsfbox{plot_dele_ppk.ps}
%\epsfxsize 1.9 truein \epsfbox{plot_mb_ppk.ps}
%}   
\centerline{
\epsfig{file=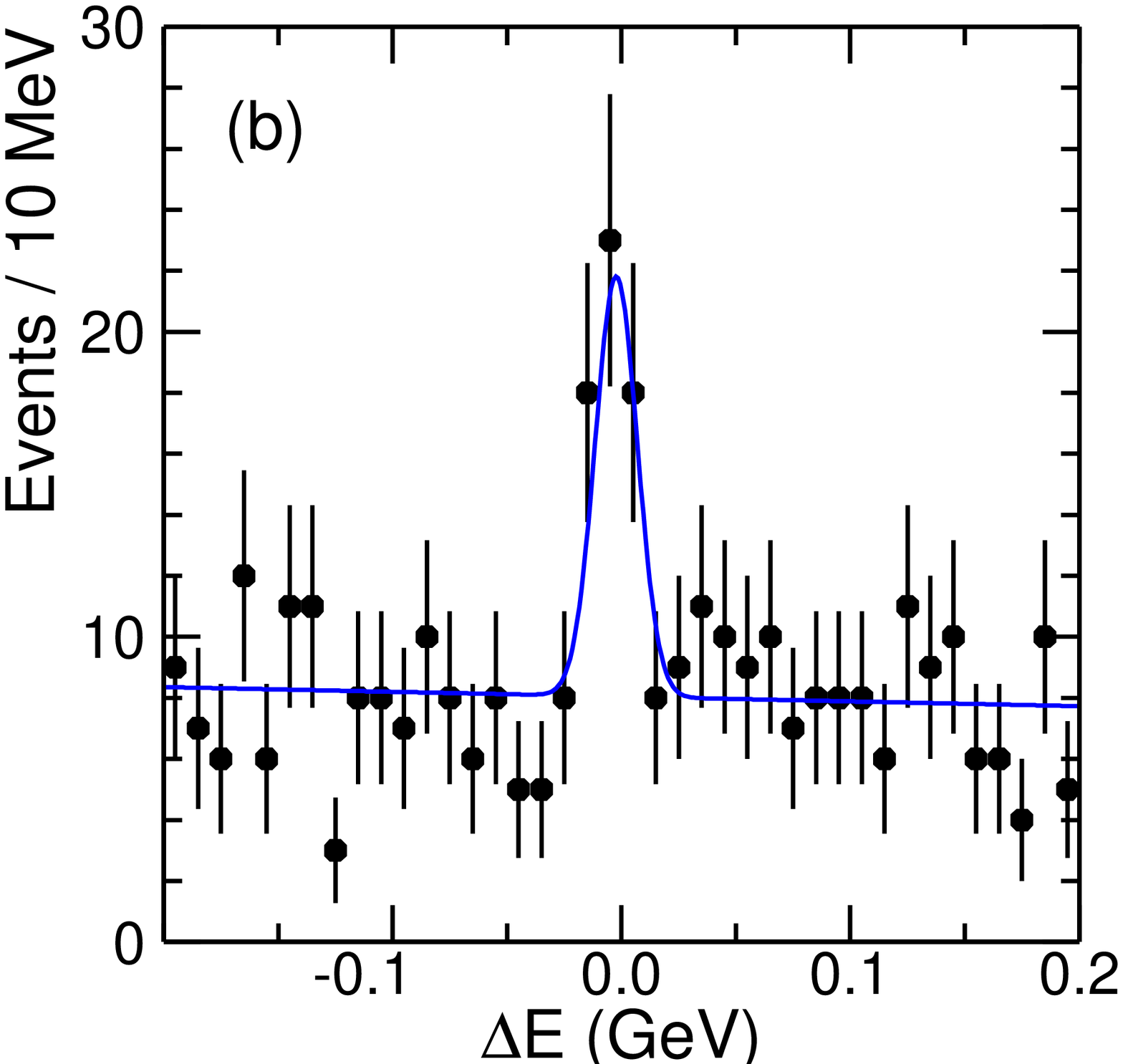,width=1.9in}
\epsfig{file=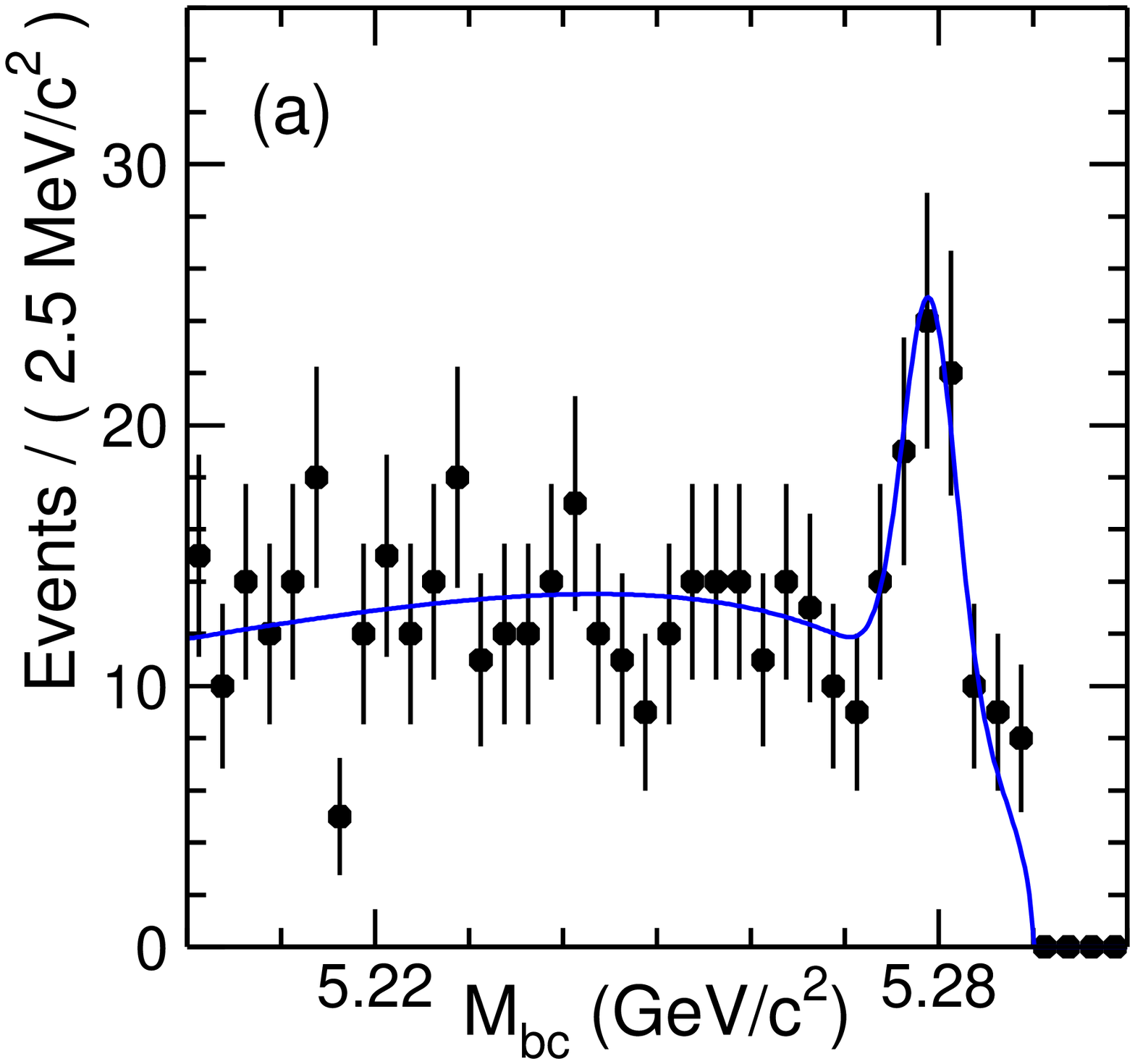,width=1.9in}
}  
\caption{(a) $\Delta E$ and (b) $M_{\rm bc}$
distributions for         
$B^+\to p\bar{p} K^+$ candidates.}
\label{mbde_ppk}
\end{figure}

In Fig.~\ref{mbde_ppk}, we show the $\Delta E$ 
and beam-constrained mass distributions for the signal candidates. 
We fit the $\Delta E$
distribution with a double Gaussian for signal 
and a linear background function with slope determined from
the $M_{\rm bc}$ sideband. The mean of the Gaussian is determined
from $\bar{B}^0\to \Lambda_c \bar{p}\pi^+\pi^-$, 
$\Lambda_c\to p K^- \pi^+$ decays.
The fit to the $\Delta E$ distribution gives a yield
of $42.8^{+10.8}_{-9.6}$ with a significance of $5.6$ standard deviations.
 In the fit to the $\Delta E$ distribution,
the region with $\Delta E<-120$ MeV is excluded to avoid feed-downs
from modes such as $B \to p \bar{p} K^{*}$. As a consistency
check, we fit the
$M_{\rm bc}$ distribution to the sum of a signal Gaussian and a background
function with kinematic threshold. The width of the
Gaussian is fixed from MC simulation while the mean
is determined from $B^+\to \bar{D}^0\pi^+$ data. The shape parameter of 
the background function is determined from $\Delta E$ sideband
data. In the $M_{\rm bc}$ distribution, 
we observe a signal of $42.9^{+9.8}_{-9.1}$ events.
%The statistical significance of the signal from the $m_B$ fit
%is $7.9$ standard deviations ($\sigma$). 
The signal yields and the branching fractions
are determined from fits to the $\Delta E$ distribution 
rather than $M_{\rm bc}$ to minimize
possible biases from $B\bar{B}$ background which tends to
peak in $M_{\rm bc}$ but not in $\Delta E$.

The background in these modes is predominantly due to continuum events.
To check for $B\bar{B}$ backgrounds that might peak in the signal 
region, we used two large $B\bar{B}$ MC samples that correspond to
an integrated luminosity that is about twice the size of
the data sample. The estimated background is of order one event
and no backgrounds that peak in the $\Delta E$ 
signal region were found.
We also examined dedicated MC
samples of $b\to c$ decay modes with baryons in the
final state. We restricted our attention to low multiplicity decay
modes. We generated samples of $\bar{B^0}\to \Lambda_c^+\bar{p}$, 
$B^-\to \Lambda_c^+ \bar{p} \pi^-$ and
$B^-\to \Lambda_c^+ \bar{p} e^-\bar{\nu}_e$
that correspond to an integrated 
luminosity about a factor of ten larger than 
the data sample used here. The $\Lambda_c$ charmed baryon
was allowed to decay into all measured decay modes
that contain a proton. Again no peaking backgrounds were observed. 

\begin{figure}[!htb]
%\centerline{
%\epsfxsize 2.8 truein \epsfbox{plot_mppbar.ps}
%}
\begin{center}
\epsfig{file=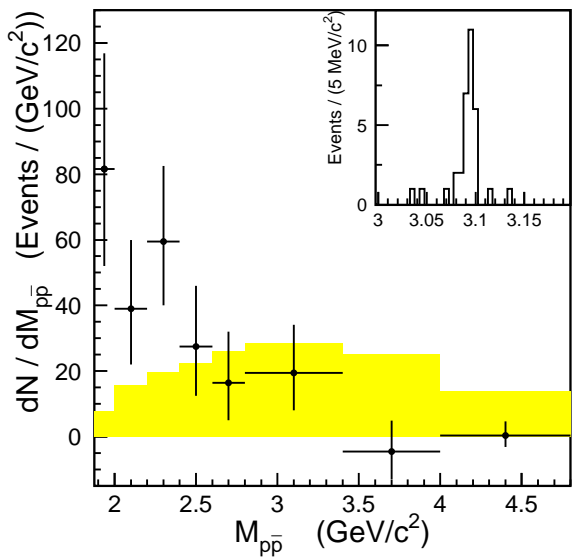, width=3.4in}      
\end{center}
\caption{The fitted yield divided by the bin size for 
$B^+\to p\bar{p} K^+$ as a function of $p \bar{p} $ mass. 
The charm veto is applied. 
The distribution from non-resonant $B^+\to p \bar{p} K^+$ 
MC simulation is superimposed. The inset shows the $p \bar{p}$ mass
distribution for the $J/\psi K^+$ signal region.}
\label{mppbar}
\end{figure}

We also examine the $M(p \bar{p})$ mass distributions for events
in the $\Delta E, ~M_{\rm bc}$ signal region. The signal yield as a function
of $p\bar{p}$ mass is shown in Fig.~\ref{mppbar}. These yields were
determined by fits to the $\Delta E$ distribution in bins of $p \bar{p}$
invariant mass. The distribution from a three-body phase space 
MC normalized to the area of
the signal is superimposed. It is clear that
the observed mass distribution is not consistent with three-body
phase space but instead is peaked at low $p\bar{p}$ mass.
We also examine the $p K^-$ mass distribution but do not
observe any obvious narrow structures such as the $\Lambda(1520)$.

To avoid model dependence in the determination of the branching
fraction for $p \bar{p} K^+$, we fit the $\Delta E$ signal yield in 
bins of $M(p \bar{p})$
and correct for the detection efficiency in each bin
using a three-body phase space
$B^+\to p \bar{p} K^+$ MC model. 
The results of the fits are given in Table I\ref{brbins}. 
We then sum the partial branching
fractions in each bin to obtain
$${\cal B}(B^+\to p \bar{p} K^+) =(4.3^{+1.1}_{-0.9}({\rm stat})\pm 0.5({\rm
syst}))\times 10^{-6}.$$
For $M(p\bar{p})<3.4$ GeV/$c^2$, the mass region below the $\chi_c$ and
$\psi^{\prime}$ resonances, 
$ {\cal B}(B^+\to p \bar{p} K^+) =(4.4^{+1.0}_{-0.8}({\rm stat})\pm
0.5({\rm syst}))\times 10^{-6}$ with the charm veto applied. For
$M(p \bar{p})<2.8$ GeV/$c^2$, the region below charm threshold, we obtain
$ {\cal B}(B^+\to p \bar{p} K^+) =(3.9^{+0.9}_{-0.7}({\rm stat})\pm
0.4({\rm syst}))\times 10^{-6}$.

\begin{table}[htb]
\label{brbins}
\begin{center}
\caption{Fit results in bins of $M(p \bar{p})$. The detection efficiency 
($\epsilon_{detect}$) and the partial branching fraction (${\cal B}$) for 
each bin are also listed.}
\begin{tabular}{c|c|c|c}
$M(p \bar{p})$(GeV/$c^2$) &  $\Delta E$ yield & $\epsilon_{detect}$ &
 ${\cal B}(\times 10^{-6})$ \\
\hline
$<2.0$ & $10.2^{+4.4}_{-3.7}$ & 0.33 &  $0.97^{+0.42}_{-0.35}$ \\ 
2.0-2.2 & $7.8^{+4.2}_{-3.4}$ & 0.34 & 
 $0.73^{+0.39}_{-0.32}$ \\ 
2.2-2.4 & $11.9^{+4.6}_{-3.9}$ & 0.30 & 
 $1.24^{+0.48}_{-0.41}$ \\ 
2.4-2.6 & $5.5^{+3.7}_{-3.0}$ & 0.29 & 
 $0.61^{+0.41}_{-0.33}$ \\ 
2.6-2.8 & $3.3^{+3.1}_{-2.3}$ & 0.30 & 
  $0.34^{+0.32}_{-0.24}$ \\ 
2.8-3.4 & $4.6^{+3.5}_{-2.7}$ & 0.29 & 
 $0.50^{+0.38}_{-0.29}$ \\
3.4-4.0 & $-1.2^{+2.5}_{-2.2}$ & 0.27 & 
 $-0.14^{+0.29}_{-0.25}$\\
4.0-4.8 & $0.3^{+3.5}_{-2.8}$ & 0.25 & 
 $0.04^{+0.45}_{-0.36}$ \\ 
\end{tabular}
\end{center}
\end{table}

% systematic errors

The contributions to the systematic error for 
the $B^+\to p \bar{p} K^+$ mode
are the uncertainties due to the tracking
efficiency (6\%), particle identification efficiency (8\%) 
and the modeling of the likelihood ratio cut (2.6\%).
The particle identification systematic includes
contributions of 3\% for the proton and anti-proton and 2\%
for the charged kaon. The error in proton/anti-proton
identification is determined using
$\Lambda/\bar{\Lambda}$ samples, while the error in kaon identification 
efficiency is obtained from kinematically selected $D^{*+}\to
D^0\pi^+, D^0\to K^-\pi^+$ in the data. The systematic
error due to the modeling of the likelihood ratio cut
is determined using $B^+\to \bar{D}^0\pi^+$ events reconstructed in data.
The systematic error in the yield of the $\Delta E$ fit (3.8\%) 
was determined by varying the mean and $\sigma$ of the signal and the shape 
parameter of the background. The sources of systematic error
are combined in quadrature to obtain the final systematic error
of 11.0\%.

For events in the $\Delta E$, $M_{\rm bc}$ signal region
we examine the proton, anti-proton and kaon particle identification
likelihood distributions and compare to signal MC simulation.
No discrepancy is observed. We also verify that the ECL shower
width distribution is consistent with MC expectations for the proton
and anti-proton candidates. In addition, we check the branching
fraction as the cuts on the proton and anti-proton probabilities
and likelihood ratio are varied. We do not observe any systematic
trends beyond statistics.

To verify the analysis procedure and branching fraction
determination, 
we remove the $J/\psi$ veto and examine the decay chain
$B^+\to J/\psi K^+$, $J/\psi\to p \bar{p}$. A clear signal of 
$26.4\pm 5.2$ events is then observed
in the $\Delta E$ spectrum.  We also observe $25.9\pm 5.1$ events
in the $M_{\rm bc}$ distribution.
The $p\bar{p}$ invariant mass spectrum for $J/\psi K^+$ 
signal candidates is shown as an inset in Fig.~\ref{mppbar}.  
We use the $\Delta E$ yield and the MC detection efficiency of 0.30
to determine the branching 
fraction ${\cal B}(B^+\to J/\psi K^+) = (13.1 \pm 2.6)\times
10^{-4}$. 
This is in good agreement with 
the PDG world average, ${\cal B}(B^+\to J/\psi K^+) = 
(10.0 \pm 1.0)\times 10^{-4}$\cite{PDG},
which was obtained by experiments
that reconstruct the $J/\psi$ in dilepton modes. 

We also examined two related decay modes 
 $B^0\to p \bar{p} K_S$ and $B^+\to p \bar{p}\pi^+$ 
that may help clarify the interpretation of the signal. 
Measurement of $B^0\to p \bar{p} K_S$
will help to determine the role of the spectator
quark in $b\to s$ decays with baryons, while observation of
$B^+\to p\bar{p}\pi^+$  will constrain the ratio of the $b\to u$ tree
 and $b\to s$ penguin diagrams in decays with baryons.
 
For  $B^0\to p \bar{p} K_S$, after the
application of the charm and $\Lambda_c$ vetoes,
no significant signal is observed in either the 
$\Delta E$ or $M_{\rm bc}$ distribution.
A fit to the $\Delta E$ distribution 
%shown in Fig.~\ref{other_modes}
gives $6.4^{+4.4}_{-3.7}$ events.
%A fit to $M_{\rm bc}$ yields $4.0\pm 3.1$ signal events.
Applying the Feldman-Cousins procedure\cite{feldman},
%In this case, the total
%number of events in the signal region is 16, and if the
%background is taken to be yy (the fit gives $9.9\pm 1.4$) then 
we obtain an upper limit of less than 16 events 
at the 90\% confidence level (C.L.). After reducing the detection
efficiency by the systematic error, we obtain an upper limit at 90\%
 C.L. of ${\cal B}(B^0\to p \bar{p} K^0)<7.2 \times 10^{-6}$. 

In the $B^+\to p \bar{p}\pi^+$ mode, after the application
of the charm veto we perform a fit to the $\Delta E$ distribution
that allows for $B^+\to p \bar{p} \pi^+$ 
signal and a reflection from misidentified
$B^+\to p \bar{p} K^+$ decays. This fit gives a 
signal yield of $16.2^{+8.6}_{-8.0}$ events and a significance of $2.1\sigma$.
% with $M(p \bar{p})<3.4$ GeV/$c^2$ 
%(see Fig.~\ref{other_modes}).
The excess in the $\Delta E$ fit corresponds to
a branching fraction ${\cal B}(B^+\to p \bar{p} \pi^+)=
(1.9^{+1.0}_{-0.9}\pm 0.3)\times 10^{-6}$ or an upper limit of 
${\cal B}(B^+\to p \bar{p} \pi^+)<3.7\times 10^{-6}$ at 90\%
C.L. after taking into account the systematic error.

We have observed a significant signal (5.6$\sigma$) for
the decay $B^+\to p \bar{p} K^+$. This is the first $b\to s$ decay
mode with baryons in the final state. In the future, this mode can
be used to search for direct $CP$ violation\cite{rosner_baryon}.
We find that
its $p \bar{p}$ mass spectrum is inconsistent with phase space
and is peaked toward low mass. This feature is suggestive of
quasi two-body decay. It is also possible that the decay is a
genuine three-body process and that this feature of the $M(p \bar{p})$
spectrum is a baryon form factor effect\cite{hou_baryon,cheng_baryon}.

%Acknowledgements

We wish to thank the KEKB accelerator group for the excellent
operation of the KEKB accelerator.
We acknowledge support from the Ministry of Education,
Culture, Sports, Science, and Technology of Japan
and the Japan Society for the Promotion of Science;
the Australian Research Council
and the Australian Department of Industry, Science and Resources;
the National Science Foundation of China under contract No.~10175071;
the Department of Science and Technology of India;
the BK21 program of the Ministry of Education of Korea
and the CHEP SRC program of the Korea Science and Engineering Foundation;
the Polish State Committee for Scientific Research
under contract No.~2P03B 17017;
the Ministry of Science and Technology of the Russian Federation;
the Ministry of Education, Science and Sport of Slovenia;
the National Science Council and the Ministry of Education of Taiwan;
and the U.S.\ Department of Energy.


\begin{thebibliography}{99}
\bibitem{CC} Hereafter, the inclusion 
of the charge conjugate mode is implied.
\bibitem{vetos} The regions  $2.85<M(p \bar{p})<3.128$ GeV/$c^2$
and $3.315<M(p \bar{p})<3.735$ GeV/$c^2$ are excluded to remove background 
from modes with $\eta_c, J/\psi$ and 
$\psi^{\prime},\chi_{c0},\chi_{c1}$ mesons, respectively.
\end{thebibliography}
\end{document}